\PassOptionsToPackage{unicode}{hyperref}
\PassOptionsToPackage{dvipsnames,svgnames*,x11names*}{xcolor}

\documentclass[]{article}

\usepackage{arxiv}

\usepackage[utf8]{inputenc} 
\usepackage[T1]{fontenc}    
\usepackage{lmodern}        
\usepackage{xcolor}
\usepackage{hyperref}       
\usepackage{url}            
\usepackage{booktabs}       
\usepackage{amsfonts}       
\usepackage{nicefrac}       
\usepackage{microtype}      
\usepackage{graphicx}
\usepackage{setspace}
\usepackage{orcidlink}
\usepackage{fontspec}

\hypersetup{
  pdftitle={On the estimation of the median odds ratio for measuring contextual effects in multilevel binary data from complex survey designs},
  pdfkeywords={multilevel data, contextual effects, median odds ratio, Delta method, MICS data},
  breaklinks=true,
  bookmarks=true,
  colorlinks=true,
  linkcolor={magenta},
  filecolor={Maroon},
  citecolor={blue},
  urlcolor={Blue},
  pdfcreator={LaTeX via pandoc}}

\title{On the estimation of the median odds ratio for measuring contextual effects in
multilevel binary data from complex survey designs}

\author{
    Shafayet Khan Shafee%
  \hspace{0.12em}\orcidlink{0009-0002-8021-3788}%
  \thanks{corresponding author}%
  \\
  Institute of Statistical Research and Training \\
  University of Dhaka \\
  Dhaka-1000, Bangladesh \\
  \texttt{\href{mailto:sshafee@isrt.ac.bd}{\nolinkurl{sshafee@isrt.ac.bd}}} \And
  M. Shafiqur Rahman%
  \hspace{0.12em}\orcidlink{0000-0001-5256-7453}%
  \\
  Institute of Statistical Research and Training \\
  University of Dhaka \\
  Dhaka-1000, Bangladesh \\
  \texttt{\href{mailto:shafiq@isrt.ac.bd}{\nolinkurl{shafiq@isrt.ac.bd}}}}

\usepackage[round,sort,compress]{natbib}

\usepackage{float}
\usepackage{caption}
\DeclareCaptionFont{tenpt}{\fontsize{9.5}{11}\selectfont}
\usepackage{bm}
\usepackage{mathtools}
\usepackage{booktabs}
\usepackage{array}
\usepackage{doi}
\usepackage{url}
\usepackage{multirow}
\usepackage{threeparttable}
\begin{document}
\maketitle

\begin{abstract}
In studies with clustered or hierarchical data structures, quantifying
between-cluster heterogeneity, referred to as contextual effects, is crucial for
valid cluster-level inference. The median odds ratio (MOR), derived from random
effects (RE) logistic regression models for clustered binary data, provides an
intuitive assessment of contextual effects. Most existing research focuses on
point estimation of the MOR for two-level models, with limited exploration of
its statistical properties under complex multilevel structures. However, the
development of corresponding interval estimators is essential for statistical
inference. Moreover, many real-world datasets, particularly those from multistage
surveys, involve hierarchical structures beyond two levels, where contextual
effects at each level are of interest. This paper discusses the estimation of
MOR for both the two-and three-level binary data, with particular emphasis on
interval estimation. Since the MOR is a post-estimation measure based on variance
components of the RE logit model, its confidence interval is derived using the
Delta method, treating the log-transformed MOR as asymptotically normal. The
approach is demonstrated across different model specifications in two-and
three-level settings. An extensive simulation study evaluated the performance of
the MOR estimators across diverse scenarios in hierarchical data settings. The
results showed that the estimators exhibited negligible bias and satisfactory
coverage probability of a 95\% confidence interval for moderate to large samples,
with small-sample bias mainly due to variance component estimation. An
application of the methods for estimating the contextual effect on C-section
delivery demonstrated that the proposed framework enhances interpretability and
supports more informed statistical and policy-oriented analyses.
\end{abstract}

\keywords{
    multilevel data
   \and
    contextual effects
   \and
    median odds ratio
   \and
    Delta method
   \and
    MICS data
  }

\setstretch{1.2}

\section{Introduction}\label{sec-01}

Multilevel or hierarchical data are nested structures in which observation units
are nested within different levels of hierarchy, producing complex variability
patterns \citep{snijders1999}. Such hierarchies may involve two levels, e.g.,
individuals within households or students within schools, and three levels, e.g.,
patients within doctors and doctors within hospitals. The hierarchical structure
of these data reflects a contextual phenomenon (or neighborhood effects), which
implies that individual behaviors and outcomes are influenced by characteristics
of the groups or environments in which they are embedded. These contextual
phenomena in multilevel data create clustering, where individuals within the
same context are not statistically independent, violating assumptions of
conventional regression \citep{steenbergen2002, hox1998}. Multilevel modeling (MLM)
addresses this issue by incorporating random effects into the model, allowing
outcome variation across levels and partitioning the total variance into cluster-
and individual-level components. The MLM not only accounts for dependency but
also enables assessment of contextual effects \citep{snijders1999, goldstein2011}.

Many socio-epidemiological and public health research emphasized that the
contextual effects in multilevel data are not a statistical nuisance but are
crucial for policy and governance \citep{merlo2001diastolic, merlo2003effect}. The
MLM quantifies the contextual effect of higher-level units (clusters) on
lower-level units using variance partition coefficient (VPC), commonly known as
intra-cluster correlation (ICC) \citep{goldstein2002}, which indicates the proportion
of total variance in the outcome attributable to the clustering context. The VPC
is straightforward to compute and interpret in multilevel linear regression
models with continuous outcomes \citep{goldstein2002}. However, for discrete responses,
calculation and interpretation are more complex because the individual-and
cluster-level variances lie on different scales. Several approaches exist to
estimate VPC, including normal response approximation, Taylor series linearization,
and the latent response method \citep{goldstein2002}. Nonetheless, limited familiarity
with latent variable concepts and assumptions such as constant individual
variance (\(\pi^2/3\)) often discourages application in epidemiology and social
science. In addition, its interpretation does not directly reflect between-cluster
heterogeneity (i.e., the clustering effects), but rather expresses the percentage
of variation due to clustering or correlation between individuals within a cluster
\citep{larsen2005}. The median odds ratio (MOR) is an alternative measure of cluster
heterogeneity, proposed by \citet{larsen2000}, which represents the median odds ratio
obtained when comparing two subjects with identical covariates but from randomly
chosen clusters, ordered by outcome odds. It quantifies the median increase in
outcome odds when moving from a lower to a higher-risk cluster and is interpreted
on odds ratio scale, similar to a fixed effect parameter \citep{larsen2005, merlo2006brief}.
\citet{austin2017} applied the concept of MOR to calculate median hazard ratio for
measuring contextual effects in multilevel survival data.

Despite the intuitive appeal of the MOR for measuring contextual effects in
multilevel data, previous studies \citep{larsen2000, larsen2005, yarnell2019} have
not addressed its interval estimation. Interval estimation is essential for
statistical inference, as it quantifies the precision and reliability of estimates,
thereby enabling more informed conclusions. Moreover, existing applications of
the MOR \citep{merlo2006brief, yarnell2019} are largely restricted to two-level data
structures, without examining its performance across different multilevel data
conditions, such as variations in outcome prevalence, sample size, number of
clusters, or cluster size. However, many surveys with complex designs generate
data with three or more hierarchical levels and varying sample sizes and outcome
prevalences. For example, the 2019 Bangladesh Multiple Indicator Cluster Survey
(MICS) \citep{mics2019} features individuals nested within households, households within
clusters (enumeration areas or small geographical regions), and clusters within
administrative districts. The survey collects data on many indicators mostly
related to maternal and child health outcomes, each exhibiting varying prevalence
rates. Therefore, estimating contextual effects at both cluster and district
levels is crucial for policy relevance in maternal and child health. Against
this background, the present study discusses the estimation of the MOR for both
two- and three-level binary data, with particular emphasis on interval estimation.
The approach is illustrated across different model specifications in multilevel
settings. An extensive simulation study is conducted to evaluate the performance
of the proposed methods under diverse multilevel data scenarios. Furthermore,
an application of the methods are demonstrated using data from the 2019 MICS to
assess the contextual effects of both small geographical areas and districts on
C-section delivery, providing some practical recommendations for researchers
studying contextual effects in multilevel binary data.

The remainder of this paper is organized as follows. Section \ref{sec-02} presents
the detailed methodology, starting with an overview of the MOR for two-level
random-intercept logistic regression model in Subsection \ref{subsec-21}, followed
by the formulation of its interval estimation in Subsection \ref{subsec-22}.
Subsections \ref{subsec-23} and \ref{subsec-24} extend the discussion to both
point and interval estimation of the MOR for two-level random-intercept-and-slope
model and three-level random-intercept model, respectively. Section \ref{sec-03}
presents the simulation study to evaluate the statistical properties of the
estimators, while Section \ref{sec-04} demonstrates a real-data application and
interpretation of the results. Finally, Section \ref{sec-05} concludes the paper
with key findings, implications, and practical recommendations.

\section{Methodology}\label{sec-02}

\subsection{MOR for two-level random-intercept logit model}\label{subsec-21}

Let us consider a two-level hierarchical structure where the \(i^{\text{th}}\)
individual (\(i = 1, 2, \ldots, n\)) is nested within the \(j^{\text{th}}\) cluster
(\(j = 1, 2, \ldots, m\)). For a set of \(p\) covariates \(\bm{x}_{ij}\), the
random-intercept logistic regression model for a binary response
\(y_{ij} \in \{0,1\}\) can be expressed as
\begin{equation}
\label{eq:two-int-logit}
\operatorname{logit}\!\left(\pi_{ij}\right) 
= \left(\beta_{0} + u_{j}\right)+ \sum_{s=1}^p {\beta}_{s}{x}_{s,ij}, \;\; 
s = 1, 2, \ldots, p,
\end{equation}
where \(\pi_{ij} = \Pr(y_{ij} = 1 \mid {x}_{s}, u)\), \({\beta}_{s}\) is the
fixed-effect regression coefficients associated with covariate \({x}_s\), and
\(u_{j} \sim \mathcal{N}(0, \sigma^{2})\) denotes the cluster-specific random
effect \citep{grilli2015}. The parameters of the model can be estimated using maximum
likelihood estimation, where adaptive Gaussian quadrature used to approximate
the likelihood \citep{goldstein1996, rodriguez2001, lesaffre2001}.

\subsection{MOR for two-level random-intercept model}\label{subsec-22}

To define MOR, let us consider two individuals with identical covariate patterns
belonging to two different clusters \(j\) and \(j'\), with random effects \(u_j\) and
\(u_{j'}\). Following \citet{larsen2000}, the odds ratio between such individuals, ordered
by the outcome odds, can be derived from the above random-intercept model
(Eq. \ref{eq:two-int-logit}) as follows
\begin{equation*}
\exp\!\left(\lvert u_j - u_{j'} \rvert\right), \quad
u_j, u_{j'} \overset{\mathrm{iid}}{\sim} \mathcal{N}(0, \sigma^2).
\end{equation*}
This odd ratio is a random variable, and for \(z>0\), its cumulative distribution
is
\begin{equation*}
\begin{aligned}
F(z) &=\; \Pr\!\left(\exp\{\lvert u_j - u_{j'} \rvert\} \le z\right)\\
&= \Pr\!\left(\lvert u_j - u_{j'} \rvert \le \log(z)\right).
\end{aligned}
\end{equation*}
Given \(u_j - u_{j'} \sim \mathcal{N}(0, 2\sigma^2)\), it follows after some
algebra that
\begin{equation*}
F(z) = 2\Phi\!\left(\frac{\log(z)}{\sqrt{2\sigma^2}}\right) - 1,
\end{equation*}
where \(\Phi^{-1}(\cdot)\) is the inverse standard normal distribution function.
Solving \(F(z)=0.5\) yields
\begin{equation*}
z_{0.5} = \exp\!\left\{\sqrt{2\sigma^2} \; \Phi^{-1}(0.75)\right\},
\end{equation*}
and then the MOR can be defined as
\begin{equation*}
\operatorname{MOR} =
\exp\!\left\{\sqrt{2\sigma^2}\,\Phi^{-1}(0.75)\right\}.
\end{equation*}
Given the estimate of \(\sigma^2\), denoted by \(\hat{\sigma}^2\), the point
estimator of MOR can be written as
\begin{equation*}
\widehat{{\operatorname{MOR}}} =\exp\!\left\{\sqrt{2\hat{\sigma}^2}\,\Phi^{-1}(0.75)\right\}.
\end{equation*}

\subsection{Interval estimation of MOR for two-level random-intercept model}\label{subsec-23}

The estimator MOR is function of the variance component of the random effect of
the model, \(\sigma^2\), which can be estimated along with its standard error using
maximum likelihood method with numerical approximation such as Gaussian or adaptive
Gaussian quadrature \citep{rodriguez2001, lesaffre2001}. Given the functional form of
the MOR estimator, we applied the Delta method \citep{ver2012} to derive its standard
error and confidence interval. The Delta method is widely used in statistical
literature to approximate the variance for a nonlinear function of an asymptotically
normal estimator by applying a first-order Taylor series expansion, thereby
linearizing the function around the estimated parameters. Following the Delta
method, we first apply a logarithmic transformation to the MOR in order to
approximate its distribution by asymptotic normality:
\begin{equation*}
\log(\widehat{{\operatorname{MOR}}}) =\sqrt{2\hat{\sigma}^2}\,\Phi^{-1}(0.75).
\end{equation*}
Using the Delta method, the variance,
\(\operatorname{Var}\bigl(\log(\widehat{\operatorname{MOR}})\bigr)\), can be derived
as follows:
\begin{equation*}
\begin{aligned}[b]
\operatorname{Var}\!\big(\log(\widehat{\operatorname{MOR}})\big) &= 
\big(\sqrt{2}\times {\Phi}^{-1} \left( 0.75\right)\big)^2 \; 
\operatorname{Var}(\hat{\sigma}) \\
\Rightarrow \operatorname{SE}\!\big(\log(\widehat{\operatorname{MOR}})\big) &= 
\big(\sqrt{2}\times {\Phi}^{-1} \left( 0.75\right)\big) \; 
\operatorname{SE}(\hat{\sigma}).
\end{aligned}
\end{equation*}
Assuming that \(\log(\widehat{{\operatorname{MOR}}})\) approximately follows
Gaussian distribution \citep{jewell2003, bland2000}, hence it is possible to write
\(\log(\widehat{\operatorname{MOR}}) \!-\! \log(\operatorname{MOR}) \sim \mathcal{N}\!\big(0, \operatorname{Var}(\log(\widehat{\operatorname{MOR}}))\big)\) and subsequently
\begin{equation*}
\frac{\log(\widehat{\operatorname{MOR}}) - \log\left(\operatorname{MOR}\right)}
     {\operatorname{SE}\!\big(\log(\widehat{\operatorname{MOR}})\big)}
\sim \mathcal{N}(0, 1).
\end{equation*}
Therefore, the two-sided \(100(1-\alpha)\%\) confidence interval for
\(\log\left({\operatorname{MOR}}\right)\) is given by
\begin{equation*}
\log(\widehat{\operatorname{MOR}}) \pm z_\alpha \, \operatorname{SE}\!\big(\log(\widehat{\operatorname{MOR}})\big),
\end{equation*}
where \(z_\alpha\) is the \((1-\alpha/2)^{th}\) percentile of the standard Gaussian
distribution. Taking anti-logarithm, the \(100(1-\alpha)\%\) confidence interval
for the \({\operatorname{MOR}}\) is given by
\begin{equation*}
\exp\left\{\log(\widehat{\operatorname{MOR}}) \pm z_\alpha \, \operatorname{SE}\!\big(\log(\widehat{\operatorname{MOR}})\big)\right\}.
\end{equation*}

\subsection{MOR for the two-level random slope model}\label{subsec-24}

Let us extend the random-intercept model (Eq. \ref{eq:two-int-logit}) by adding
a random-slope for a single covariate, say \(x\), to account for the variation in
the effect of \(x\) across clusters:
\begin{equation}
\label{eq:two-slope-logit}
\operatorname{logit}\left( \pi_{ij}\right) = (\beta_0 + u_{1j}) + \left(\beta_1 + u_{2j}\right) x_{ij}  \; ,
\end{equation}
where \(u_{1j} \sim \mathcal{N}\left(0, \sigma_1^2\right)\) and
\(u_{2j} \sim \mathcal{N}\left(0, \sigma_2^2\right)\). Assume that
\(\operatorname{Cov}(u_{1j}, u_{2j}) = 0\). For a specific value of \(x=x_0\), the
odds ratio for two individuals from different clusters, say \(j\) and \(j'\), can be
derived from Eq. \eqref{eq:two-slope-logit} and written as
\begin{equation*}
\exp\!\bigl\{\lvert (u_{1j} - u_{1j'})+(u_{2j}-u_{2j'})x_0 \rvert\bigr\}.
\end{equation*}
Letting \(\kappa(x_0) = \lvert (u_{1j} - u_{1j}) + (u_{2j}-u_{2j'}) x_0 \rvert\),
the cumulative distribution of the odd ratio for \(z>0\) is given by
\begin{equation*}
\begin{aligned}
F(z \mid x = x_0) &=\; \Pr\bigl\{\exp(\kappa(x_0)\le z)\bigr\}\\
&= \Pr\bigl\{\kappa(x_0) \le \log(z)\bigr\}.
\end{aligned}
\end{equation*}
Given that \(\kappa(x_0)\sim \mathcal{N}\left(0, \operatorname{V}_{x_0}\right)\),
where \(\operatorname{V}_{x_0} = 2(\sigma^2_1 + x_0^2\sigma^2_2)\) for
\(\operatorname{Cov}(u_{1j}, u_{2j}) = 0\), it becomes
\begin{equation*}
F(z \mid x=x_0) = 2\Phi\!\left(\frac{\log(z)}{\sqrt{\operatorname{V}_{x_0}}}\right) - 1,
\end{equation*}
where \(\Phi^{-1}(\cdot)\) is the inverse standard normal distribution function.
Solving \(F(z \mid x=x_0)=0.5\) yields the \(\operatorname{MOR}(x_0)\) as follows:
\begin{equation*}
\operatorname{MOR}(x_0) =
\exp\!\left\{\sqrt{\operatorname{V}(x_0)}\times\Phi^{-1}(0.75)\right\}.
\end{equation*}
Given the estimate of \(\operatorname{V}(x_0)\), denoted by
\(\hat{\operatorname{V}}(x_0) = 2(\hat{\sigma}^2_1 + x_0^2\hat{\sigma}^2_2)\), the
point estimator of \(\operatorname{MOR}(x_0)\) can be written as
\begin{equation*}
\widehat{{\operatorname{MOR}}}(x_0) =\exp\!\left\{\sqrt{\hat{\operatorname{V}}(x_0)}\times\Phi^{-1}(0.75)\right\}.
\end{equation*}
Now following the same argument and methods discussed in the earlier section to
derive confidence interval for the MOR, the two-sided \(100(1-\alpha)\%\) confidence
interval for \(\widehat{MOR}(x_0)\) is given by,
\begin{equation*}
\exp\Bigl\{\log(\widehat{\operatorname{MOR}}(x_0)) \pm z_\alpha \, \operatorname{SE}\bigl(\log(\widehat{\operatorname{MOR}}(x_0))\bigr)\Bigr\} \;,
\end{equation*}
where \(z_\alpha\) is the \((1-\alpha/2)^{th}\) percentile of the standard Gaussian
distribution, and
\begin{equation*}
\begin{aligned}[b]
\operatorname{SE}\bigl(\log(\widehat{\operatorname{MOR}}(x_0))\bigr) = \Bigl\{\bm{J\Sigma J}^T\Bigr\}^{\frac{1}{2}} \, ,
\end{aligned}
\end{equation*}
with
\begin{equation*}
\bm{\Sigma} = \begin{bmatrix}
\operatorname{Var}(\hat{\sigma}_1) &  0\\
0 & \operatorname{Var}(\hat{\sigma}_2)
\end{bmatrix} \, ,
\end{equation*}
and the Jacobian matrix
\begin{equation*}
\bm{J}=\begin{bmatrix}
\frac{\partial \log(\widehat{\operatorname{MOR}}(x_0))}{\partial \hat{\sigma}_1} & \frac{\partial \log(\widehat{\operatorname{MOR}}(x_0))}{\partial \hat{\sigma}_2}
\end{bmatrix} \, .
\end{equation*}

\subsection{MOR for the three-level random-intercept model}\label{subsec-25}

Let us consider a three-level data scenario where \(i^{th}\) individual is nested
within the \(j^{th}\) unit, which is itself nested within \(k^{th}\) unit, with
\(i=1,2,\dots,n\), \(j=1,2,\dots,m\) and \(k=1,2,\dots,q\). The model in
Eq. \eqref{eq:two-int-logit} can be extended to a three-level random-intercept
model as follows
\begin{equation}
\label{eq:three-int-logit}
\operatorname{logit}\left( \pi_{ijk}\right) = \beta_0  + u_{jk} + v_{k} +\sum_{s=1}^p{\beta}_{s}{x}_{s,ijk} \;,
\end{equation}
where \(\pi_{ijk} = \Pr\!\left(y_{ijk}=1 \mid x,u,v\right)\) is the event
probability of the binary outcome \(y_{ijk}\in\{0,1\}\),
\(u_{jk} \sim \mathcal{N}\left( 0, \sigma_{u}^{2}\right)\),
\(v_{k} \sim \mathcal{N}\left(0, \sigma_{v}^{2}\right)\) and
\(\operatorname{Cov}(u_{jk}, v_{k}) = 0\).

For the three-level data scenario, two types of contextual effect and associated
MORs are to be estimated, one for the effect of the second level (\(j\)) and the
other for the effect of the third level (\(k\)). Let us denote \(\operatorname{MOR}_1\)
and \(\operatorname{MOR}_2\) for the contextual effect of level 2 and level 3,
respectively. Following the same arguments as discussed earlier, the odd ratio
between two individuals with the same covariate patterns from two different units
of level 2, say \(j\) and \(j'\), both nested within the same unit of level 3, say
\(k\), can be defined as \(\exp{\left(|u_{jk} - u_{j'k}|\right)}\). The point
estimator for the median odd ratio (\(\operatorname{MOR}_{1}\)) can be derived by
solving \(F(z)=0.5\) and written as follows
\begin{equation*}
\widehat{\operatorname{MOR}}_1 = \exp\left\{ \sqrt{2\hat{\sigma}_{u}^{2}} \times {\Phi}^{-1}\left( 0.75 \right)\right\},
\end{equation*}
where \({\Phi}^{-1}\left(\cdot\right)\) refers to the inverse standard normal
distribution function, \(\hat{\sigma}_u\) is the estimate of the variance component
of the level 2 random effects in the model (Eq. \ref{eq:three-int-logit}).
Similarly, following the Delta methods, the two-sided \(100(1-\alpha)\%\) confidence
interval for \(\widehat{\operatorname{MOR}}_1\) can be estimated as follows:
\begin{equation*}
\exp\Bigl\{\log(\widehat{\operatorname{MOR}}_1) \pm z_\alpha \times \operatorname{SE}\bigl(\log(\widehat{\operatorname{MOR}}_1)\bigr)\Bigr\} \;,
\end{equation*}
where \(z_\alpha\) is the \((1-\alpha/2)^{th}\) percentile of the standard Gaussian
distribution, and
\begin{equation*}
\begin{aligned}[b]
 \operatorname{SE}\bigl(\log(\widehat{\operatorname{MOR}}_1)\bigr) = \bigl(\sqrt{2}\times {\Phi}^{-1} \left( 0.75\right)\bigr) \; \operatorname{SE}(\hat{\sigma}_u) \; .
\end{aligned}
\end{equation*}
Again, the odd ratio between two individuals with the same covariate patterns
from two different units of level 2, say \(j\) and \(j'\) from two different units
of level 3, say \(k\) and \(k'\), can be defined as
\(\exp\left\{{\lvert(u_{jk} - u_{j'k'}) + (v_k - v_{k'})}\rvert\right\}\). The
point estimator for the median odd ratio (\(\operatorname{MOR}_2\)) for the
subject with higher odds versus those with lower odds can be derived by
solving its cumulative distribution \(F(z)=0.5\) and written as follows:
\begin{equation*}
\widehat{\operatorname{MOR}}_2 = \exp\Bigl\{ \sqrt{2\left(\hat{\sigma}_u^2 + \hat{\sigma}_v^2\right)} \times {\Phi}^{-1}\left( 0.75 \right)\Bigr\},
\end{equation*}
where \(\hat{\sigma}^2_u\) and \(\hat{\sigma}^2_v\) are the estimated variance
components of the random effects associated the level 2 and 3, respectively. Now,
similarly as before, the two-sided \(100(1-\alpha)\%\) confidence interval for \(\widehat{\operatorname{MOR}}_2\) is given by,
\begin{equation*}
\exp\Bigl\{\log(\widehat{\operatorname{MOR}}_2) \pm z_\alpha \times \operatorname{SE}\bigl(\log(\widehat{\operatorname{MOR}}_2)\bigr)\Bigr\} ,
\end{equation*}
where \(z_\alpha\) is the \((1-\alpha/2)^{th}\) percentile of the standard Gaussian
distribution, and
\begin{equation*}
\begin{aligned}[b]
\operatorname{SE}\bigl(\log(\widehat{\operatorname{MOR}}_2)\bigr) = \Bigl\{\bm{J}_1\bm{\Sigma}_1 \bm{J}_1^T\Bigr\}^{\frac{1}{2}} \;,
\end{aligned}
\end{equation*}
with
\begin{equation*}
\bm{\Sigma}_1 = \begin{bmatrix}
\operatorname{Var}(\hat{\sigma}_u) &  0\\
0 & \operatorname{Var}(\hat{\sigma}_v)
\end{bmatrix},
\end{equation*}
and the Jacobian matrix,
\begin{equation*}
\bm{J}_1=\begin{bmatrix}
\frac{\partial \log(\widehat{\operatorname{MOR}}_2)}{\partial \hat{\sigma}_u} & \frac{\partial \log(\widehat{\operatorname{MOR}}_2)}{\partial \hat{\sigma}_v}
\end{bmatrix}.
\end{equation*}

\section{Simulation study}\label{sec-03}

An extensive simulation study was conducted to evaluate the performance of both
point and interval estimators of the MOR derived from different forms of
multilevel logistic regression models. The aim is to assess the statistical
properties of the estimators under diverse scenarios of multilevel data and to
provide practical recommendations.

\subsection{Simulation scenarios}\label{subsec-31}

Following the methodological discussion for MOR, three forms of multilevel
models were considered: i) two-level random-intercept model, ii) two-level
random-slope model, iii) three-level random-intercept model. For each model
type, several simulation scenarios were considered under a factorial simulation
design by varying the outcome prevalence rate, number of clusters and the
cluster size. For all models, two levels of outcome prevalence were examined:
low (prevalence: 10\% on average), and moderate (prevalence: 30\% on average).
For the two-level models (both random-intercept-and-slope), the number of
clusters varied as \(m\in\{10, 30, 50, 10\}\) and the cluster size as
\(n\in \{5, 10, 30, 50\}\). For the random-slope model, three additional scenarios
were included by fixing the value of the continuous covariate, associated with
the random-slope, at its three quartile points. For three-level random-intercept
model, the number of units varied as follows, level 1: \(q\in \{20, 40, 100\}\),
level 2: \(m\in\{10, 30\}\), and level 3: \(n\in\{5, 15, 30\}\). Combining all
scenarios, there were 32 simulation scenarios for each of the two-level models
(total of 64 scenarios) and 36 for three-level models. For each scenario, a
total of 1000 replications were considered.

\subsection{Data generation}\label{subsec-32}

For each simulation setting of the two-and three-level data structures,
covariates, random effects, and outcomes were generated from a true model
corresponding to the specified scenario and the type of random-effects model for
which the MOR was to be estimated.

\subsection*{Two-level random-intercept model}\label{two-level-random-intercept-model}
\addcontentsline{toc}{subsection}{Two-level random-intercept model}

For the random-intercept model with a two-level structure comprising \(m\)
clusters and \(n\) individuals per cluster (a total of \(m \times n\) observations),
two continuous covariates were generated independently from \(\mathcal N(0,1)\),
with one treated as continuous, denoted by \(x_c\) and the other dichotomized at
\(0.5\) to form a binary covariate, denoted by \(x_b\). Cluster-specific random
effects \(u_j \; (j=1, \ldots, m)\) were generated from \(\mathcal N(0,2.5)\) and
assigned to all \(n\) individuals within the cluster. The binary responses were
then generated from Bernoulli distribution with event probability
\(\pi_{ij} = \exp(\eta_{ij})/(1+\exp(\eta_{ij}))\), where the linear predictor for
the true model was specified as
\begin{equation*}
\eta_{ij} = \beta_0 + \beta_c x_{c,ij} + \beta_b x_{b,ij} + u_j ,
\end{equation*}
with \(\beta_c=1.75\), \(\beta_b=0.67\). The value of the intercept as fixed as
\(\beta_0=-1.85\) and \(-4.1\) to target prevalence-level of the outcome as
moderated (prevalence: 30\% on average) or low (prevalence: 10\% on average),
respectively.

\subsection*{Two-level random-slope model}\label{two-level-random-slope-model}
\addcontentsline{toc}{subsection}{Two-level random-slope model}

Keeping the same covariate generation as before, for the random-slope
specification, cluster-specific random effects
\((u_{1j}, u_{2j})^\top \sim \mathcal N_2(\bm{0},\bm{D})\), with
\begin{equation*}
\bm{D} = \begin{bmatrix} 1 & 0 \\ 0 & 2 \end{bmatrix},
\end{equation*}
were generated and assigned to all \(n\) individuals within the cluster. The
responses \(y_{ij}\) were generated from \(\text{Bernoulli}(\pi_{ij})\), where
\(\pi_{ij}\) was defined analogously using the following linear predictor:
\begin{equation*}
\text{logit}({\pi_{ij}})=\eta_{ij} = \beta_0 + u_{1j} + (\beta_c+u_{2j})x_{c,ij} + \beta_b x_{b,ij},
\end{equation*}
with parameter values for \((\beta_0,\beta_c,\beta_b)\) as above.

\subsection*{Three-level random-intercept model}\label{three-level-random-intercept-model}
\addcontentsline{toc}{subsection}{Three-level random-intercept model}

For a three-level structure with \(n\) units at level 1 nested within \(m\) units at
level 2, and \(q\) units at level 3 (\(i=1, \dots, n; j=1,\ldots, m; k=1,\ldots, q\)),
resulting in a total of \(N = q \times m \times n\) observations, covariates were
generated as before. Random effects at level 2, \(u_{jk} \sim \mathcal N(0,2)\),
and at level 3, \(v_k \sim \mathcal N(0,2.5)\), were added to the random-intercept
model. The responses \(y_{ijk}\) were then generated from \(\text{Bernoulli}(\pi_{ijk})\),
where \(\pi_{ijk}\) was defined with the linear predictor:
\begin{equation*}
\text{logit}(\pi_{ijk})=\eta_{ijk} = \beta_0 + \beta_c x_{c,ijk} + \beta_b x_{b,ijk} + u_{jk} + v_k,
\end{equation*}
with the same parameterization of \(\beta_0,\beta_c,\beta_b\) to control prevalence.

\subsection{Metrics to evaluate the performance of MOR and software used for estimating the parameters}\label{subsec-33}

For each dataset in a simulation scenario, the desired model was fitted using
the standard maximum likelihood procedure with adaptive Gaussian quadrature.
Based on the parameters of the fitted model, the MOR was computed for each
dataset, and the average over 1000 valid estimates (excluding those from simulated
datasets where the likelihood estimation failed to converge) is reported. To
evaluate the performance of the point estimator, the relative bias was computed as,
relative bias = \((\widehat{\operatorname{MOR}} - \operatorname{ MOR})/\operatorname{MOR}\),
where the true value of the MOR was obtained using the parameter values specified
in the data-generating true model. For interval estimation, the coverage
probability of the 95\% nominal confidence interval was calculated as the percentage
of intervals that contained the true MOR value across the 1000 replications.

The simulations and the associated computations were conducted using the R
programming language (\texttt{v4.3.1}). The two-level random-intercept and
random-slope models are fitted using the \texttt{mixed\_model} function of the
\texttt{GLMMadaptive} package (\texttt{v0.9-0}). The three-level random-intercept
models are fitted using the \texttt{glmmTMB} function of the \texttt{glmmTMB} package
(\texttt{v1.1.7}). In addition, several other packages, such as \texttt{merDeriv}
(\texttt{v0.2-4}), \texttt{numDeriv} (\texttt{v2016.8-1.1}), \texttt{purrr}
(\texttt{v1.0.1}), \texttt{tidyr} (\texttt{v1.3.0}), \texttt{stringr} (\texttt{v1.5.0}),
\texttt{ggplot2} (\texttt{v3.4.2}), \texttt{dplyr} (\texttt{v1.1.2}) were used
for summarizing the simulation results. All R codes used in this study can be
found in a GitHub repository \url{https://github.com/shafayetShafee/contained-thesis}.

\subsection{Simulation results}\label{subsec-34}

\subsubsection*{MOR for the two-level random-intercept model}\label{mor-for-the-two-level-random-intercept-model}
\addcontentsline{toc}{subsubsection}{MOR for the two-level random-intercept model}

Table \ref{tab:ih2} presents the results for the case where the prevalence of
the binary responses is moderate (\(29\%\)). Since the MOR depends on the model
parameters, the performance of the maximum likelihood estimates (MLEs) of the
model parameters was first examined to assess whether the model was correctly
estimated, particularly its variance component of the random effects (\(\sigma^2\)).
The results indicate that the MLEs of the model parameters closely approximate
their true values in large samples. In particular, the regression coefficients
exhibited smaller bias with increasing cluster size. However, the performance of
the variance component estimate depended on both the number of clusters and
cluster size, showing an inverse relationship between bias and sample size.
Consequently, the performance of \(\widehat{\text{MOR}}\) followed a pattern
similar to that of \(\hat{\sigma}^2\), with relative bias decreasing as the total
sample size increased. Specifically, for a fixed number of clusters,
\(\widehat{\text{MOR}}\) tended to exhibit negative bias (i.e., underestimate the
true value) as cluster size increased.

\vspace{-1mm}

\begin{table}[!h]
\centering
\caption{\label{tab:ih2}Estimates of the parameters of the two-level random-intercept model, and estimates of MOR, relative bias (\%), coverage probability of 95\% nominal confidence interval. The estimates are reported as the average over 1000 simulations. The prevalence of outcome variable is on average 29\%. True $\beta_0 = -1.85$, $\beta_c = 1.75$, $\beta_b = 0.67$, $\sigma^2_u=2.5$, MOR = 4.52.}
\centering
\fontsize{8.5}{9.5}\selectfont
\begin{tabular}[t]{>{\centering\arraybackslash}m{0.8cm}>{\centering\arraybackslash}m{0.8cm}>{\centering\arraybackslash}m{1cm}>{\centering\arraybackslash}m{1cm}>{\centering\arraybackslash}m{1cm}>{\centering\arraybackslash}m{1cm}>{\centering\arraybackslash}m{1cm}>{\centering\arraybackslash}m{1.3cm}>{\centering\arraybackslash}m{1.4cm}}
\toprule
m & n & $\hat{\beta}_0$ & $\hat{\beta}_1$ & $\hat{\beta}_2$ & $\hat{\sigma}_u^2$ & $\widehat{\text{MOR}}$ & Rel. Bias (\%) & 95\% CI Coverage\\
\midrule
 & 5 & -2.05 & 2.03 & 0.70 & 3.10 & 6.41 & 41.80 & 0.94\\

 & 10 & -1.92 & 1.88 & 0.69 & 2.82 & 5.50 & 21.65 & 0.94\\

 & 30 & -1.90 & 1.80 & 0.68 & 2.47 & 4.67 & 3.43 & 0.88\\

\multirow{-4}{0.8cm}{\centering\arraybackslash 10} & 50 & -1.88 & 1.78 & 0.70 & 2.30 & 4.37 & -3.36 & 0.86\\
\cmidrule{1-9}
 & 5 & -1.94 & 1.86 & 0.67 & 2.87 & 5.42 & 19.86 & 0.97\\

 & 10 & -1.88 & 1.80 & 0.67 & 2.58 & 4.74 & 4.88 & 0.94\\

 & 30 & -1.86 & 1.76 & 0.68 & 2.48 & 4.54 & 0.50 & 0.92\\

\multirow{-4}{0.8cm}{\centering\arraybackslash 30} & 50 & -1.87 & 1.76 & 0.68 & 2.45 & 4.49 & -0.57 & 0.91\\
\cmidrule{1-9}
 & 5 & -1.90 & 1.82 & 0.67 & 2.76 & 5.08 & 12.48 & 0.96\\

 & 10 & -1.89 & 1.79 & 0.68 & 2.60 & 4.72 & 4.41 & 0.94\\

 & 30 & -1.86 & 1.76 & 0.67 & 2.48 & 4.52 & -0.04 & 0.93\\

\multirow{-4}{0.8cm}{\centering\arraybackslash 50} & 50 & -1.87 & 1.76 & 0.67 & 2.46 & 4.48 & -0.78 & 0.95\\
\cmidrule{1-9}
 & 5 & -1.87 & 1.76 & 0.66 & 2.61 & 4.74 & 4.83 & 0.94\\

 & 10 & -1.87 & 1.76 & 0.67 & 2.50 & 4.54 & 0.56 & 0.94\\

 & 30 & -1.86 & 1.75 & 0.67 & 2.46 & 4.48 & -0.85 & 0.94\\

\multirow{-4}{0.8cm}{\centering\arraybackslash 100} & 50 & -1.86 & 1.75 & 0.67 & 2.48 & 4.50 & -0.37 & 0.95\\
\bottomrule
\multicolumn{9}{l}{\rule{0pt}{1em}$m=$ number of cluster, $n=$ cluster size, Rel. Bias = Relative Bias, CI = Confidence interval}\\
\end{tabular}
\end{table}

\vspace{-1mm}

A similar trend was observed for the interval estimator of MOR, which almost
achieved coverage of the 95\% nominal confidence interval under large sample
conditions. Comparable patterns of the results in the estimation of model
parameters and the MOR were observed when the outcome prevalence is low (10\%)
(Supplementary Table \ref{tab:il2}). However, the bias was slightly higher for
the corresponding small-sample scenarios compared to those with prevalence of 29\%.

\subsubsection*{MOR for the two-level random-slope model}\label{mor-for-the-two-level-random-slope-model}
\addcontentsline{toc}{subsubsection}{MOR for the two-level random-slope model}

The results in Table \ref{tab:sh2} summarize the performance of the MOR estimators
under a two-level random-slope logistic regression model with a moderate outcome
prevalence (27\%). Overall, across all quartiles of the covariate associated with
the random slope, the MOR estimator demonstrates robust performance, with reduced
bias and appropriate coverage as sample size increases. In particular, the bias
tends to decrease as both the number of clusters (\(m\)) and cluster size (\(n\))
increase, indicating improved estimator stability with larger samples. For small
cluster sizes (\(n=10\)) and few clusters (\(m=10\)), the MOR is slightly underestimated,
reflecting downward (negative) bias due to underestimation of the variance component.
The coverage probabilities of the 95\% confidence intervals are generally close
to the nominal level (approximately 0.95), for large sample conditions, suggesting
consistency in the estimation. A similar pattern of results was observed when
outcome prevalence was low (9\%), however, the bias was slightly higher in the
corresponding small sample scenarios (Supplementary Table \ref{tab:sl2}).

\subsubsection*{MOR for the three-level random-intercept model}\label{mor-for-the-three-level-random-intercept-model}
\addcontentsline{toc}{subsubsection}{MOR for the three-level random-intercept model}

Table 3 presents the results of MOR from three-level random intercept model with
moderate outcome prevalence (31\%). The estimates of the MOR at both the second
and third levels indicate that the proposed estimators (\(\widehat{\text{MOR}}_1\)
and \(\widehat{\text{MOR}}_2\)) perform well as the number of level-2 and level-3
units increases, reflecting asymptotically unbiased estimation with larger sample
sizes. Relative bias is generally small (within \(\pm7\%\)), and it decreases
systematically with increasing cluster sizes. Coverage probabilities of the
nominal 95\% confidence intervals remain close to the expected level (0.95) for
moderate to large samples but tend to fall below the nominal level in small-sample
scenarios (e.g., \(n=5\) or \(m=10\)), particularly for the third-level MOR. This is
because small samples at higher levels tend to produce slightly underestimated
variance components. Similar pattern of findings were observed for the case with
a low outcome prevalence (12\%), see Supplementary Table \ref{tab:il3}.

\vspace{-1mm}

\begin{table}[H]
\centering
\caption{\label{tab:sh2}Estimates of MOR, relative bias (\%), coverage probability of 95\% nominal confidence interval. The MOR was computed at three quartiles of the covariate associated with random slope. The estimates are reported as the average over 1000 simulations. The prevalence of outcome variable is on average 27\%.}
\centering
\fontsize{8}{9}\selectfont
\begin{tabular}[t]{>{\centering\arraybackslash}m{0.6cm}>{\centering\arraybackslash}m{0.6cm}>{\centering\arraybackslash}m{1.4cm}>{\centering\arraybackslash}m{1.4cm}>{\centering\arraybackslash}m{1.4cm}>{\centering\arraybackslash}m{1.4cm}>{\centering\arraybackslash}m{1.4cm}>{\centering\arraybackslash}m{1.4cm}}
\toprule
\multicolumn{1}{c}{ } & \multicolumn{1}{c}{ } & \multicolumn{2}{c}{$Q_{1X_{1c}}$} & \multicolumn{2}{c}{$Q_{2X_{1c}}$} & \multicolumn{2}{c}{$Q_{3X_{1c}}$} \\
\cmidrule(l{3pt}r{3pt}){3-4} \cmidrule(l{3pt}r{3pt}){5-6} \cmidrule(l{3pt}r{3pt}){7-8}
m & n & Rel. Bias (\%) & 95\% CI Coverage & Rel. Bias (\%) & 95\% CI Coverage & Rel. Bias (\%) & 95\% CI Coverage\\
\midrule
 & 5 & 23.87 & 0.99 & 15.73 & 0.99 & 34.39 & 1.00\\

 & 10 & 20.44 & 0.99 & 9.46 & 0.96 & 24.79 & 0.98\\

 & 30 & 0.33 & 0.96 & -1.24 & 0.90 & 3.44 & 0.96\\

\multirow{-4}{0.6cm}{\centering\arraybackslash 10} & 50 & -2.50 & 0.95 & -3.39 & 0.89 & -1.84 & 0.96\\
\cmidrule{1-8}
 & 5 & 21.81 & 0.97 & 12.39 & 0.98 & 24.52 & 0.98\\

 & 10 & 3.99 & 0.97 & 0.84 & 0.97 & 8.50 & 0.97\\

 & 30 & -0.47 & 0.93 & -1.41 & 0.92 & -0.36 & 0.97\\

\multirow{-4}{0.6cm}{\centering\arraybackslash 30} & 50 & -1.76 & 0.93 & -1.55 & 0.93 & -1.06 & 0.96\\
\cmidrule{1-8}
 & 5 & 13.77 & 0.96 & 7.72 & 0.97 & 18.80 & 0.96\\

 & 10 & 2.22 & 0.95 & -0.01 & 0.96 & 4.34 & 0.97\\

 & 30 & -1.06 & 0.92 & -0.82 & 0.94 & 0.27 & 0.97\\

\multirow{-4}{0.6cm}{\centering\arraybackslash 50} & 50 & -1.32 & 0.94 & -0.86 & 0.93 & -0.50 & 0.96\\
\cmidrule{1-8}
 & 5 & 3.24 & 0.96 & 2.95 & 0.97 & 7.47 & 0.96\\

 & 10 & 1.17 & 0.94 & 0.45 & 0.96 & 2.14 & 0.96\\

 & 30 & -0.28 & 0.91 & 0.00 & 0.92 & 0.45 & 0.95\\

\multirow{-4}{0.6cm}{\centering\arraybackslash 100} & 50 & -0.87 & 0.93 & -0.14 & 0.94 & 0.29 & 0.96\\
\bottomrule
\multicolumn{8}{l}{\rule{0pt}{1em}$m=$ number of cluster, $n=$ cluster size, Rel. Bias = Relative Bias, CI = Confidence interval}\\
\end{tabular}
\end{table}

\vspace{-2mm}

\begin{table}[!h]
\centering
\caption{\label{tab:ih3}Point estimate, relative bias and the coverage of the 95\% nominal confidence interval for both second ($\text{MOR}_1$) and third level MOR ($\text{MOR}_2$) from the three-level random intercept logistic model. The prevalence of outcome variable is on average 31\%. The estimates are reported as the average over 1000 simulations.}
\centering
\fontsize{8}{9}\selectfont
\begin{tabular}[t]{>{\centering\arraybackslash}m{0.6cm}>{\centering\arraybackslash}m{0.6cm}>{\centering\arraybackslash}m{0.6cm}>{\centering\arraybackslash}m{1.2cm}>{\centering\arraybackslash}m{1.2cm}>{\centering\arraybackslash}m{1.2cm}>{\centering\arraybackslash}m{1.2cm}>{\centering\arraybackslash}m{1.2cm}>{\centering\arraybackslash}m{1.2cm}}
\toprule
\multicolumn{1}{c}{ } & \multicolumn{1}{c}{ } & \multicolumn{1}{c}{ } & \multicolumn{3}{c}{Second Level} & \multicolumn{3}{c}{Third Level} \\
\cmidrule(l{3pt}r{3pt}){4-6} \cmidrule(l{3pt}r{3pt}){7-9}
q & m & n & $\widehat{\text{MOR}}$ & Rel. Bias (\%) & 95\% CI Coverage & $\widehat{\text{MOR}}$ & Rel. Bias (\%) & 95\% CI Coverage\\
\midrule
   & 10 & 5 & 3.64 & -5.65 & 0.90 & 7.13 & -5.72 & 0.88\\
20 & 10 & 15 & 3.79 & -1.77 & 0.93 & 7.25 & -4.21 & 0.88\\
   & 10 & 30 & 3.83 & -0.56 & 0.93 & 7.42 & -1.90 & 0.89\\
\midrule
   & 30 & 5 & 3.58 & -7.02 & 0.85 & 6.98 & -7.74 & 0.83\\
20 & 30 & 15 & 3.78 & -2.01 & 0.93 & 7.29 & -3.61 & 0.88\\
   & 30 & 30 & 3.83 & -0.74 & 0.95 & 7.37 & -2.52 & 0.89\\
\midrule
   & 10 & 5 & 3.59 & -6.76 & 0.91 & 7.03 & -7.00 & 0.85\\
40 & 10 & 15 & 3.80 & -1.51 & 0.94 & 7.44 & -1.70 & 0.91\\
   & 10 & 30 & 3.83 & -0.58 & 0.95 & 7.45 & -1.45 & 0.91\\
\midrule
   & 30 & 5 & 3.57 & -7.23 & 0.78 & 6.97 & -7.92 & 0.84\\
40 & 30 & 15 & 3.78 & -1.92 & 0.92 & 7.28 & -3.76 & 0.90\\
   & 30 & 30 & 3.82 & -0.86 & 0.95 & 7.41 & -2.01 & 0.92\\
\bottomrule
\multicolumn{9}{l}{\rule{0pt}{1em}$q$ = number of units in level 3, $m$ = number units in level 2, $n$ = number of units in level 1,}\\
\multicolumn{9}{l}{\rule{0pt}{1em}Rel. Bias = Relative Bias, CI = Confidence interval}\\
\end{tabular}
\end{table}

\vspace{-1mm}

\section{Application: contextual effect on C-section delivery}\label{sec-04}

\subsection{Data and variables}\label{data-and-variables}

An application of the proposed methods is demonstrated by estimating the MOR to
assess the contextual effects of both enumeration areas (cluster: primary sampling
units) and administrative districts on C-section (caesarean) delivery among
women of reproductive age (15--49 years). Among women who gave birth in health
facilities in recent years, about 68\% delivered by C-section. This rate is high
and has been rising rapidly over the past two decades, and researchers and
policymakers consider it a major public-health concern. Therefore, evaluating
the contextual influences of small geographical areas and administrative
districts C-section delivery is crucial for evidence-based policy formulation
targeting regions with higher or lower rates.

Data used in this study were extracted from the 2019 Bangladesh Multiple
Indicator Cluster Survey (MICS) \citep{mics2019}. The MICS is a nationally
representative household survey that employed a two-stage stratified cluster
sampling scheme to select households and the respondents, considering sixty-four
administrative districts as strata (domains). In the first stage, enumeration
areas (EAs) were selected systematically from the population census frame with
probabilities proportional to stratum size, resulting in a total of 3,220 EAs
across 64 strata. In the second stage, 20 households were systematically selected
from each sampled EA, yielding a final sample of 64,400 households. From each
household, one woman of reproductive age (15--49 years) was interviewed. The present
analysis is based on 4813 women who gave birth in facilities in the two years
preceding the survey, and the outcome of interest is whether a woman delivered
birth by C-section or not (1=yes, 0=no).

A set of covariates reflecting socio-demographic characteristics of women and
their households were considered: age, education (no education or pre-primary,
primary, secondary and higher secondary+), household economic status (poorest,
poorer, middle, richer, richest), received antenatal care (yes, no), type of
healthcare facilities (public, private), watch television (yes, no).

\subsection{Exploring and estimating contextual effects}\label{exploring-and-estimating-contextual-effects}

\begin{figure}
\centering
\includegraphics[width=0.7\textwidth]{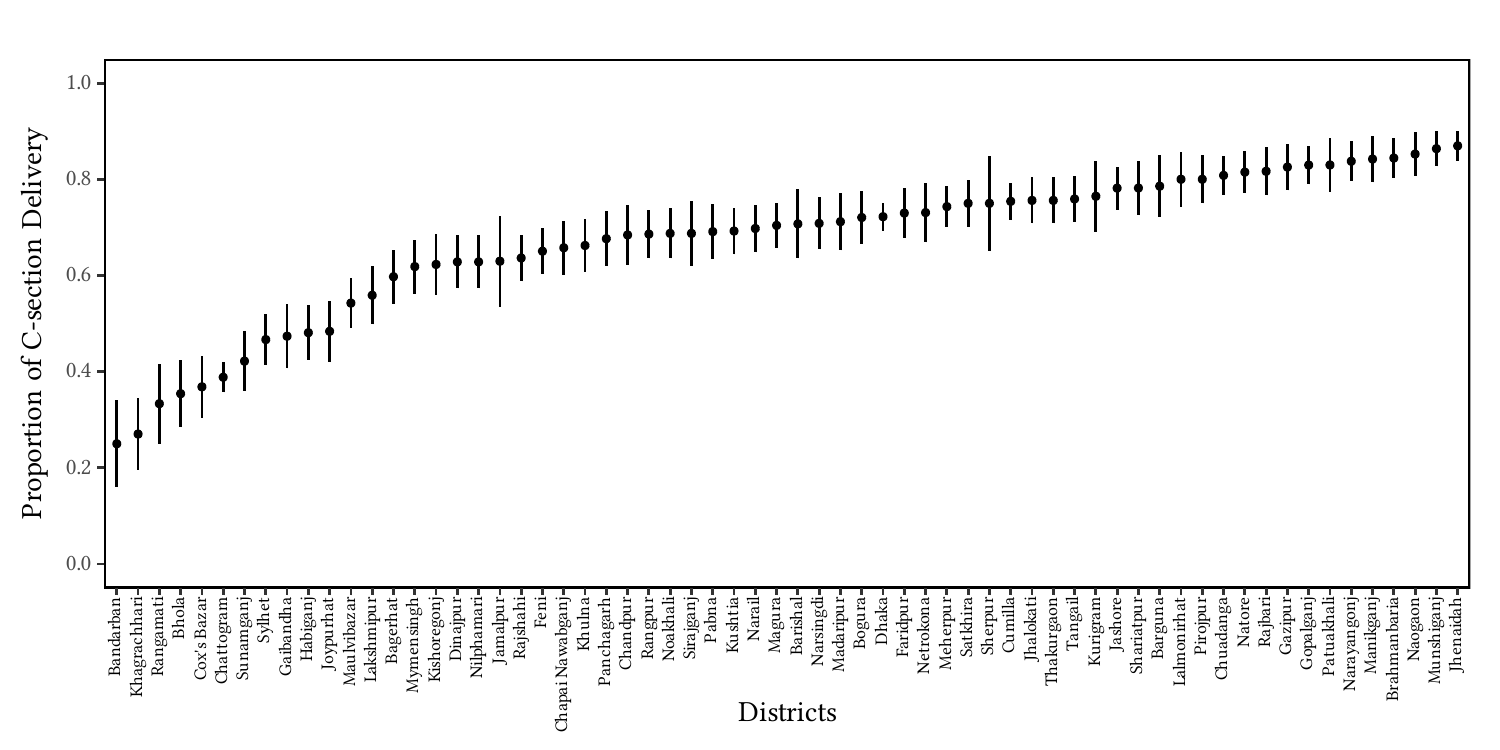}
\caption{District-wise proportion of women who gave birth with C-section in a 
health facilities (overall use rate = 0.68), along with their confidence 
intervals, displayed in the ranked order of districts.}
\label{fig:dist-prop}
\end{figure}

Under the sampling design in MICS 2019, households or respondents represent the
first-level units nested within enumeration areas (EAs) at the second level,
which in turn are nested within districts at the third level. To explore the
contextual effects of administrative districts on the C-section delivery, the
district-specific proportions of women with C-section delivery were ranked and
plotted along with standard error (confidence interval plots) for 64 districts
(Figure \ref{fig:dist-prop}).

\vspace{-0.8cm}

The graphs revealed substantial variation in C-section rates across administrative
districts, indicating notable contextual effects at the district level. Similar
patterns were observed across EAs (results not shown). These findings motivated
the quantification of contextual effects at both the EAs and district levels
through estimation of the MOR.

Given the presence of contextual effects at the second and third levels of
clustering, three models were formulated to estimate the contextual variation
in C-section delivery at each level. First, a two-level random-intercept logistic
model was fitted, allowing the variation across enumeration areas (EAs) while
ignoring districts. Second, the model was extended by incorporating a random-slope
for women's age, allowing both the intercepts (between-EA variation) and the slopes
(age-related variation within EAs) to vary randomly across EAs. The MOR was then
estimated at the three quartiles of the age variable. The third model was a
three-level random-intercept model, accounting for variation in C-section delivery
attributable to EAs at level 2 and to districts at level 3. For both the two- and
three-level random-intercept models, estimation was performed twice: first without
covariates (intercept only) and then with a set of individual-level covariates
to examine how covariate-adjustment affected the estimated contextual effects.

The MOR estimates (Table \ref{tab:mor}) from the two-level random-intercept model
indicate substantial between-EA heterogeneity in C-section delivery, with an MOR
of 2.10 in the null model. This suggests that two women with identical individual
characteristics but living in different EAs would differ, on average, twofold
in their odds of giving C-section delivery. After adjusting for individual-level
covariates, the MOR decreased to 2.09, indicating that part of the between-EA
variation is explained by compositional differences in the individual level
covariates. With the inclusion of a random slope for age, the MORs at the first,
second, and third quartiles of age were 1.85, 2.07, and 2.33, respectively,
implying slightly stronger contextual effects among older women. In the three-level
random-intercept model, the estimated MORs were 1.51 for EAs and 2.08 for districts
in the null model, which declined to 1.49 and 1.89, respectively, after covariate
adjustment. These reductions indicate that the inclusion of individual-level
factors accounts for some, but not all, of the observed contextual variation due
to EAs and districts. The persistence of relatively high MOR values even after
adjustment suggests that unobserved contextual factors at both the EA and district
levels play a significant role in shaping women's C-section delivery in Bangladesh.

\begin{table}[!h]
\centering
\caption{\label{tab:mor}The MOR from the multilevel logit model to measure the 
contextual effects of enumeration area (EA: primary sampling unit) and 
administrative districts in the C-section delivery by the women of age 15--49 
years.}
\centering
\fontsize{9}{10}\selectfont
\begin{threeparttable}
\begin{tabular}[t]{>{\raggedright\arraybackslash}p{3.3cm}>{\raggedright\arraybackslash}p{2.8cm}>{\centering\arraybackslash}p{0.8cm}>{\centering\arraybackslash}p{1.9cm}>{\centering\arraybackslash}p{0.8cm}>{\centering\arraybackslash}p{1.9cm}}
\toprule
\multicolumn{2}{c}{ } & \multicolumn{2}{c}{Null Model} & \multicolumn{2}{c}{Adjusted Model} \\
\cmidrule(l{3pt}r{3pt}){3-4} \cmidrule(l{3pt}r{3pt}){5-6}
Multi-level model format & Type of MOR & $\widehat{\text{MOR}}$ & 95\% CI & $\widehat{\text{MOR}}$ & 95\% CI\\
\midrule
Two-level\newline random-intercept & $\text{MOR}_{\text{EA}}$ & 2.10 & (1.84, 2.39) & 2.09 & (1.76, 2.47)\\
\cmidrule{1-6}
\addlinespace[2pt]
 & $\text{MOR}_{\text{Q}_1:\,\text{age}=21}$ & -- & -- & 1.85 & (1.61, 2.12)\\

\addlinespace[2pt]
 & $\text{MOR}_{\text{Q}_2:\,\text{age}=25}$ & -- & -- & 2.07 & (1.76, 2.44)\\

\addlinespace[2pt]
\multirow{-3}{3.3cm}{\raggedright\arraybackslash Two-level\newline random-slope} & $\text{MOR}_{\text{Q}_3:\,\text{age}=29}$ & -- & -- & 2.33 & (1.93, 2.81)\\
\cmidrule{1-6}
\addlinespace[2pt]
 & $\text{MOR}_{1,\text{EA}}$ & 1.51 & (1.27, 1.78) & 1.49 & (1.17, 1.90)\\

\addlinespace[2pt]
\multirow{-2}{3.3cm}{\raggedright\arraybackslash Three-level\newline random-intercept} & $\text{MOR}_{2,\text{District}}$ & 2.08 & (1.80, 2.40) & 1.89 & (1.58, 2.27)\\
\bottomrule
\end{tabular}
\begin{tablenotes}
\item Null model: model without any covariates; Adjusted model: model with a set 
of covariates: age, women's education, education of household head, marital status, 
own a mobile phone, ever used computer or tablet, household economic status.
\end{tablenotes}
\end{threeparttable}
\end{table}

\section{Discussion}\label{sec-05}

This paper discusses the estimation of median odds ratio (MOR) for two- and
three-level random effects logistic regression models, with particular emphasis
on interval estimation. Specifically, the estimation of the MOR is demonstrated
for three types of random-effects models: a two-level model with random intercept
only, a two-level model with both random intercept and slope, and a three-level
model with random intercept only. The \(\widehat{\text{MOR}}\) is a function of
the variance components of the random effects in the models; therefore its
computation follows by the estimation of the model parameters and can be
regarded as a post-estimation procedure. Accordingly, interval estimation of the
MOR is developed using Delta method \citep{ver2012}. The \(\widehat{\text{MOR}}\)
provides an intuitive interpretation of contextual effects in terms of odds
ratios and is comparable to the interpretation of fixed effects. The performance
and key statistical properties of the proposed methods was evaluated under
diverse scenarios of multilevel data structures through an extensive simulation
study.

The simulation results indicated that both the point and interval estimators of
the MOR, derived from the all types of multilevel models under study, performed
well by providing estimates with negligible bias and satisfactory coverage
probabilities of the 95\% confidence interval, particularly when the sample
size was moderate to large. The small sample bias in \(\widehat{\text{MOR}}\) was
primarily induced by bias in the estimated variance components of the random
effects under limited sample conditions (i.e., small number of clusters, each
with small cluster-size). Overall, the simulation findings suggest that the
performance of \(\widehat{\text{MOR}}\) is comparable to that of other
maximum-likelihood-based estimators commonly reported in the statistical
literature. Since \(\widehat{\text{MOR}}\) is derived from the MLEs of the variance
components, it inherits the asymptotic properties of MLEs, such as consistency
and approximate unbiasedness. However, to reduce small-sample bias in variance
component of the random effects of the model, alternatives to standard MLE,
such as Bayesian approach with integrated nested Laplace approximation \citep{fong2010}
or MLE with higher number of quadrature points in adaptive Gaussian
approximation \citep{rodriguez1995, rodriguez2001, lesaffre2001} can be applied to
estimate the model parameters.

Furthermore, the proposed methods are illustrated using data from a nationally
representative cluster survey in Bangladesh (MICS 2019) to quantify the contextual
effects of small regional groups (enumeration areas) and administrative districts
on C-section delivery in health facilities among women of reproductive age
(15--49 years). The estimated MOR showed an intuitive and policy-relevant
interpretation of the contextual influences. The results revealed substantial
and persistent contextual disparities in C-section delivery across both
enumeration areas and districts, even after accounting for individual-level
characteristics. These findings suggest that, beyond personal attributes,
structural and community level factors play a critical role in shaping C-section
delivery. This illustration offers comprehensive guidelines for researchers to
apply the proposed methods to measure the contextual effects on other outcomes
in MICS 2019 and similar complex surveys, thereby supporting the development of
evidence-based policy recommendations.

In conclusion, the proposed framework for estimating the MOR with confidence
intervals enhances the Interpretability of contextual effects in the random-effects
logistic models for multilevel binary data, particularly from cluster survey with
complex designs, thereby supporting more informed statistical inference and
policy-oriented analysis in hierarchical data settings.

\section*{Acknowledgments}\label{acknowledgments}
\addcontentsline{toc}{section}{Acknowledgments}

\vspace{-1em}

The authors acknowledge the UNICEF and Bangladesh Bureau of Statistics (BBS) for
providing access the survey data used in this study.

\section*{Data availability}\label{data-availability}
\addcontentsline{toc}{section}{Data availability}

\vspace{-1em}

The dataset used in this research are accessible to a public domain
here: \url{https://mics.unicef.org/surveys}

\section*{Competing interests}\label{competing-interests}
\addcontentsline{toc}{section}{Competing interests}

\vspace{-1em}

The authors declared none.

\section*{Funding}\label{funding}
\addcontentsline{toc}{section}{Funding}

\vspace{-1em}

This research did not receive any specific grant from funding agencies in the
public, commercial, or not-for-profit sectors.

\section*{ORCID}\label{orcid}
\addcontentsline{toc}{section}{ORCID}

\vspace{-1em}

Shafayet Khan Shafee \orcidlink{0009-0002-8021-3788} \url{https://orcid.org/0009-0002-8021-3788}

M. Shafiqur Rahman \orcidlink{0000-0001-5256-7453} \url{https://orcid.org/0000-0001-5256-7453}

\renewcommand\refname{References}

\newpage
\section*{Supplementary Materials}

\setcounter{table}{0}
\renewcommand{\thetable}{S\arabic{table}}

\begin{table}[H]
\centering
\caption{\label{tab:il2}Estimates of the parameters of the two-level random-intercept model, and estimates of MOR, relative bias (\%), coverage probability of 95\% nominal confidence interval. The estimates are reported as the average over 1000 simulations. The prevalence of outcome variable is on average 10\%. True $\beta_0 = -4.1$, $\beta_c = 1.75$, $\beta_b = 0.67$, $\sigma^2_u=2.5$, MOR = 4.52.}
\centering
\fontsize{9}{10}\selectfont
\begin{tabular}[t]{>{\centering\arraybackslash}m{0.8cm}>{\centering\arraybackslash}m{0.8cm}>{\centering\arraybackslash}m{1cm}>{\centering\arraybackslash}m{1cm}>{\centering\arraybackslash}m{1cm}>{\centering\arraybackslash}m{1cm}>{\centering\arraybackslash}m{1cm}>{\centering\arraybackslash}m{1.3cm}>{\centering\arraybackslash}m{1.4cm}}
\toprule
m & n & $\hat{\beta}_0$ & $\hat{\beta}_1$ & $\hat{\beta}_2$ & $\hat{\sigma}_u^2$ & $\widehat{\text{MOR}}$ & Rel. Bias (\%) & 95\% CI Coverage\\
\midrule
 & 5 & -4.09 & 1.94 & 0.77 & 2.19 & 4.84 & 7.08 & 0.91\\

 & 10 & -4.30 & 1.93 & 0.70 & 2.67 & 5.48 & 21.26 & 0.92\\

 & 30 & -4.24 & 1.82 & 0.67 & 2.50 & 4.82 & 6.59 & 0.89\\

\multirow{-4}{0.8cm}{\centering\arraybackslash 10} & 50 & -4.17 & 1.79 & 0.66 & 2.34 & 4.53 & 0.24 & 0.86\\
\cmidrule{1-9}
 & 5 & -4.37 & 1.92 & 0.64 & 3.02 & 6.01 & 32.91 & 0.97\\

 & 10 & -4.24 & 1.81 & 0.69 & 2.64 & 4.94 & 9.28 & 0.97\\

 & 30 & -4.14 & 1.77 & 0.65 & 2.52 & 4.63 & 2.54 & 0.91\\

\multirow{-4}{0.8cm}{\centering\arraybackslash 30} & 50 & -4.13 & 1.77 & 0.65 & 2.46 & 4.52 & -0.04 & 0.94\\
\cmidrule{1-9}
 & 5 & -4.29 & 1.87 & 0.67 & 2.88 & 5.56 & 23.12 & 0.99\\

 & 10 & -4.19 & 1.79 & 0.69 & 2.65 & 4.88 & 7.92 & 0.96\\

 & 30 & -4.11 & 1.77 & 0.66 & 2.47 & 4.53 & 0.24 & 0.93\\

\multirow{-4}{0.8cm}{\centering\arraybackslash 50} & 50 & -4.12 & 1.76 & 0.67 & 2.47 & 4.51 & -0.11 & 0.95\\
\cmidrule{1-9}
 & 5 & -4.19 & 1.80 & 0.69 & 2.67 & 4.92 & 8.87 & 0.98\\

 & 10 & -4.14 & 1.77 & 0.66 & 2.54 & 4.64 & 2.62 & 0.94\\

 & 30 & -4.11 & 1.76 & 0.67 & 2.47 & 4.49 & -0.58 & 0.94\\

\multirow{-4}{0.8cm}{\centering\arraybackslash 100} & 50 & -4.12 & 1.75 & 0.67 & 2.47 & 4.49 & -0.60 & 0.93\\
\bottomrule
\multicolumn{9}{l}{\rule{0pt}{1em}$m=$ number of cluster, $n=$ cluster size, Rel. Bias = Relative Bias, CI = Confidence interval}\\
\end{tabular}
\end{table}

\begin{table}[!h]
\centering
\caption{\label{tab:sl2}Estimates of MOR, relative bias (\%), coverage probability of 95\% nominal confidence interval. The MOR was computed at three quartiles of the covariate associated with random slope. The estimates are reported as the average over 1000 simulations. The prevalence of outcome variable is on average 9\%.}
\centering
\fontsize{9}{10}\selectfont
\begin{tabular}[t]{>{\centering\arraybackslash}m{0.6cm}>{\centering\arraybackslash}m{0.6cm}>{\centering\arraybackslash}m{1.4cm}>{\centering\arraybackslash}m{1.4cm}>{\centering\arraybackslash}m{1.4cm}>{\centering\arraybackslash}m{1.4cm}>{\centering\arraybackslash}m{1.4cm}>{\centering\arraybackslash}m{1.4cm}}
\toprule
\multicolumn{1}{c}{ } & \multicolumn{1}{c}{ } & \multicolumn{2}{c}{$Q_{1X_{1c}}$} & \multicolumn{2}{c}{$Q_{2X_{1c}}$} & \multicolumn{2}{c}{$Q_{3X_{1c}}$} \\
\cmidrule(l{3pt}r{3pt}){3-4} \cmidrule(l{3pt}r{3pt}){5-6} \cmidrule(l{3pt}r{3pt}){7-8}
m & n & Rel. Bias (\%) & 95\% CI Coverage & Rel. Bias (\%) & 95\% CI Coverage & Rel. Bias (\%) & 95\% CI Coverage\\
\midrule
 & 5 & -9.23 & 1.00 & 4.68 & 1.00 & 8.90 & 1.00\\

 & 10 & 2.75 & 1.00 & 5.34 & 1.00 & 17.20 & 1.00\\

 & 30 & 4.29 & 0.99 & 0.82 & 0.98 & 5.42 & 0.99\\

\multirow{-4}{0.6cm}{\centering\arraybackslash 10} & 50 & 1.22 & 0.99 & -2.37 & 0.95 & -1.20 & 0.98\\
\cmidrule{1-8}
 & 5 & 18.55 & 1.00 & 18.64 & 1.00 & 33.84 & 1.00\\

 & 10 & 15.39 & 0.98 & 10.76 & 1.00 & 21.05 & 0.98\\

 & 30 & -3.15 & 0.98 & -4.26 & 0.95 & -0.27 & 0.99\\

\multirow{-4}{0.6cm}{\centering\arraybackslash 30} & 50 & -4.96 & 0.97 & -3.66 & 0.96 & -0.63 & 0.98\\
\cmidrule{1-8}
 & 5 & 26.16 & 0.99 & 16.42 & 1.00 & 27.40 & 0.99\\

 & 10 & 9.96 & 0.97 & 3.39 & 0.98 & 10.40 & 0.98\\

 & 30 & -3.14 & 0.96 & -2.76 & 0.96 & 0.98 & 0.98\\

\multirow{-4}{0.6cm}{\centering\arraybackslash 50} & 50 & -1.76 & 0.94 & -1.45 & 0.95 & 0.24 & 0.98\\
\cmidrule{1-8}
 & 5 & 14.75 & 0.97 & 8.76 & 0.99 & 16.60 & 0.98\\

 & 10 & -0.09 & 0.96 & -1.71 & 0.96 & 1.64 & 0.98\\

 & 30 & -1.64 & 0.91 & -1.79 & 0.96 & -0.77 & 0.99\\

\multirow{-4}{0.6cm}{\centering\arraybackslash 100} & 50 & -1.03 & 0.91 & -0.90 & 0.95 & 0.06 & 0.98\\
\bottomrule
\multicolumn{8}{l}{\rule{0pt}{1em}$m=$ number of cluster, $n=$ cluster size, Rel. Bias = Relative Bias, CI = Confidence interval}\\
\end{tabular}
\end{table}

\begin{table}[!h]
\centering
\caption{\label{tab:il3}Point estimate, relative bias and the coverage of the 95\% nominal confidence interval for both second ($\text{MOR}_1$) and third level MOR ($\text{MOR}_2$) from the three-level random intercept logistic model. The prevalence of outcome variable is on average 12\%. The estimates are reported as the average over 1000 simulations.}
\centering
\fontsize{9}{10}\selectfont
\begin{tabular}[t]{>{\centering\arraybackslash}m{0.6cm}>{\centering\arraybackslash}m{0.6cm}>{\centering\arraybackslash}m{0.6cm}>{\centering\arraybackslash}m{1.2cm}>{\centering\arraybackslash}m{1.2cm}>{\centering\arraybackslash}m{1.4cm}>{\centering\arraybackslash}m{1.2cm}>{\centering\arraybackslash}m{1.2cm}>{\centering\arraybackslash}m{1.4cm}}
\toprule
\multicolumn{1}{c}{ } & \multicolumn{1}{c}{ } & \multicolumn{1}{c}{ } & \multicolumn{3}{c}{Second Level} & \multicolumn{3}{c}{Third Level} \\
\cmidrule(l{3pt}r{3pt}){4-6} \cmidrule(l{3pt}r{3pt}){7-9}
q & m & n & $\widehat{\text{MOR}}$ & Rel. Bias (\%) & 95\% CI Coverage & $\widehat{\text{MOR}}$ & Rel. Bias (\%) & 95\% CI Coverage\\
\midrule
   & 10 & 5 & 3.85 & -0.18 & 0.94 & 7.19 & -5.00 & 0.89\\
20 & 10 & 15 & 3.70 & -3.99 & 0.91 & 7.04 & -6.97 & 0.86\\
   & 10 & 30 & 3.79 & -1.77 & 0.93 & 7.26 & -3.97 & 0.89\\
\midrule
   & 30 & 5 & 3.51 & -9.04 & 0.86 & 6.68 & -11.66 & 0.80\\
20 & 30 & 15 & 3.68 & -4.38 & 0.88 & 7.04 & -6.95 & 0.85\\
   & 30 & 30 & 3.76 & -2.33 & 0.92 & 7.21 & -4.65 & 0.87\\
\midrule
   & 10 & 5 & 3.66 & -5.01 & 0.90 & 6.91 & -8.66 & 0.87\\
40 & 10 & 15 & 3.72 & -3.58 & 0.92 & 7.19 & -4.95 & 0.86\\
   & 10 & 30 & 3.77 & -2.08 & 0.93 & 7.32 & -3.26 & 0.89\\
\midrule
   & 30 & 5 & 3.46 & -10.17 & 0.78 & 6.57 & -13.10 & 0.75\\
40 & 30 & 15 & 3.70 & -3.93 & 0.88 & 7.05 & -6.75 & 0.84\\
   & 30 & 30 & 3.76 & -2.51 & 0.89 & 7.24 & -4.27 & 0.89\\
\bottomrule
\multicolumn{9}{l}{\rule{0pt}{1em}$q$ = number of units in level 3, $m$ = number units in level 2, $n$ = number of units in level 1,}\\
\multicolumn{9}{l}{\rule{0pt}{1em}Rel. Bias = Relative Bias, CI = Confidence interval}\\
\end{tabular}
\end{table}


\begin{thebibliography}{}

\bibitem[Austin et~al., 2017]{austin2017}
Austin, P.~C., Wagner, P., and Merlo, J. (2017).
\newblock The median hazard ratio: a useful measure of variance and general contextual effects in multilevel survival analysis.
\newblock {\em Statistics in Medicine}, 36(6):928--938.
\newblock \href {https://doi.org/10.1002/sim.7188} {\path{DOI: 10.1002/sim.7188}}.

\bibitem[BBS and UNICEF, 2019]{mics2019}
BBS and UNICEF (2019).
\newblock {\em Progotir Pathey, Bangladesh Multiple Indicator Cluster Survey 2019, Survey Findings Report}.
\newblock Bangladesh Bureau of Statistics (BBS) and UNICEF Bangladesh, Dhaka, Bangladesh.

\bibitem[Bland and Altman, 2000]{bland2000}
Bland, J.~M. and Altman, D.~G. (2000).
\newblock The odds ratio.
\newblock {\em BMJ}, 320(7247):1468.
\newblock \href {https://doi.org/10.1136/bmj.320.7247.1468} {\path{DOI: 10.1136/bmj.320.7247.1468}}.

\bibitem[Fong et~al., 2010]{fong2010}
Fong, Y., Rue, H., and Wakefield, J. (2010).
\newblock Bayesian inference for generalized linear mixed models.
\newblock {\em Biostatistics}, 11(3):397--412.
\newblock \href {https://doi.org/10.1093/biostatistics/kxp053} {\path{DOI: 10.1093/biostatistics/kxp053}}.

\bibitem[Goldstein, 2011]{goldstein2011}
Goldstein, H. (2011).
\newblock {\em Multilevel Statistical Models}.
\newblock Wiley Series in Probability and Statistics. John Wiley \& Sons Ltd, 4th edition.
\newblock \href {https://doi.org/10.1002/9780470973394} {\path{DOI: 10.1002/9780470973394}}.

\bibitem[Goldstein et~al., 2002]{goldstein2002}
Goldstein, H., Browne, W., and Rasbash, J. (2002).
\newblock Partitioning variation in multilevel models.
\newblock {\em Understanding Statistics}, 1(4):223--231.
\newblock \href {https://doi.org/10.1207/S15328031US0104\_02} {\path{DOI: 10.1207/S15328031US0104\_02}}.

\bibitem[Goldstein and Rasbash, 1996]{goldstein1996}
Goldstein, H. and Rasbash, J. (1996).
\newblock Improved approximations for multilevel models with binary responses.
\newblock {\em Journal of the Royal Statistical Society Series A: Statistics in Society}, 159(3):505--513.
\newblock \href {https://doi.org/10.2307/2983328} {\path{DOI: 10.2307/2983328}}.

\bibitem[Grilli and Rampichini, 2015]{grilli2015}
Grilli, L. and Rampichini, C. (2015).
\newblock Specification of random effects in multilevel models: a review.
\newblock {\em Quality {\&} Quantity}, 49(3):967--976.
\newblock \href {https://doi.org/10.1007/s11135-014-0060-5} {\path{DOI: 10.1007/s11135-014-0060-5}}.

\bibitem[Hoef, 2012]{ver2012}
Hoef, J. M.~V. (2012).
\newblock Who invented the delta method?
\newblock {\em The American Statistician}, 66(2):124--127.
\newblock \href {https://doi.org/10.1080/00031305.2012.687494} {\path{DOI: 10.1080/00031305.2012.687494}}.

\bibitem[Hox, 1998]{hox1998}
Hox, J. (1998).
\newblock {Multilevel Modeling: When and Why}.
\newblock In Balderjahn, I., Mathar, R., and Schader, M., editors, {\em Classification, Data Analysis, and Data Highways}, pages 147--154, Berlin, Heidelberg. Springer Berlin Heidelberg.
\newblock \href {https://doi.org/10.1007/978-3-642-72087-1_17} {\path{DOI: 10.1007/978-3-642-72087-1_17}}.

\bibitem[Jewell, 2003]{jewell2003}
Jewell, N.~P. (2003).
\newblock {\em {Statistics for Epidemiology}}.
\newblock Chapman \& Hall/CRC Texts in Statistical Science. CRC press, 1st edition.

\bibitem[Larsen and Merlo, 2005]{larsen2005}
Larsen, K. and Merlo, J. (2005).
\newblock Appropriate assessment of neighborhood effects on individual health: Integrating random and fixed effects in multilevel logistic regression.
\newblock {\em American Journal of Epidemiology}, 161(1):81--88.
\newblock \href {https://doi.org/10.1093/aje/kwi017} {\path{DOI: 10.1093/aje/kwi017}}.

\bibitem[Larsen et~al., 2000]{larsen2000}
Larsen, K., Petersen, J.~H., Budtz-Jørgensen, E., and Endahl, L. (2000).
\newblock Interpreting parameters in the logistic regression model with random effects.
\newblock {\em Biometrics}, 56(3):909--914.
\newblock \href {https://doi.org/10.1111/j.0006-341X.2000.00909.x} {\path{DOI: 10.1111/j.0006-341X.2000.00909.x}}.

\bibitem[Lesaffre and Spiessens, 2001]{lesaffre2001}
Lesaffre, E. and Spiessens, B. (2001).
\newblock On the effect of the number of quadrature points in a logistic random effects model: An example.
\newblock {\em Journal of the Royal Statistical Society Series C: Applied Statistics}, 50(3):325--335.
\newblock \href {https://doi.org/10.1111/1467-9876.00237} {\path{DOI: 10.1111/1467-9876.00237}}.

\bibitem[Merlo et~al., 2006]{merlo2006brief}
Merlo, J., Chaix, B., Ohlsson, H., Beckman, A., Johnell, K., Hjerpe, P., R{\r a}stam, L., and Larsen, K. (2006).
\newblock A brief conceptual tutorial of multilevel analysis in social epidemiology: using measures of clustering in multilevel logistic regression to investigate contextual phenomena.
\newblock {\em Journal of Epidemiology \& Community Health}, 60(4):290--297.
\newblock \href {https://doi.org/10.1136/jech.2004.029454} {\path{DOI: 10.1136/jech.2004.029454}}.

\bibitem[Merlo et~al., 2003]{merlo2003effect}
Merlo, J., Lynch, J.~W., Yang, M., Lindström, M., Östergren, P.~O., Rasmusen, N.~K., and Råstam, L. (2003).
\newblock Effect of neighborhood social participation on individual use of hormone replacement therapy and antihypertensive medication: A multilevel analysis.
\newblock {\em American Journal of Epidemiology}, 157(9):774--783.
\newblock \href {https://doi.org/10.1093/aje/kwg053} {\path{DOI: 10.1093/aje/kwg053}}.

\bibitem[Merlo et~al., 2001]{merlo2001diastolic}
Merlo, J., {\"O}stergren, P.-O., Hagberg, O., Lindstr{\"o}m, M., Lindgren, A., Melander, A., R{\r a}stam, L., and Berglund, G. (2001).
\newblock Diastolic blood pressure and area of residence: multilevel versus ecological analysis of social inequity.
\newblock {\em Journal of Epidemiology \& Community Health}, 55(11):791--798.
\newblock \href {https://doi.org/10.1136/jech.55.11.791} {\path{DOI: 10.1136/jech.55.11.791}}.

\bibitem[Rodríguez and Goldman, 1995]{rodriguez1995}
Rodríguez, G. and Goldman, N. (1995).
\newblock An assessment of estimation procedures for multilevel models with binary responses.
\newblock {\em Journal of the Royal Statistical Society Series A: Statistics in Society}, 158(1):73--89.
\newblock \href {https://doi.org/10.2307/2983404} {\path{DOI: 10.2307/2983404}}.

\bibitem[Rodríguez and Goldman, 2001]{rodriguez2001}
Rodríguez, G. and Goldman, N. (2001).
\newblock Improved estimation procedures for multilevel models with binary response: A case-study.
\newblock {\em Journal of the Royal Statistical Society Series A: Statistics in Society}, 164(2):339--355.
\newblock \href {https://doi.org/10.1111/1467-985X.00206} {\path{DOI: 10.1111/1467-985X.00206}}.

\bibitem[Snijders and Bosker, 2012]{snijders1999}
Snijders, T. A.~B. and Bosker, R.~J. (2012).
\newblock {\em Multilevel Analysis: An Introduction to Basic and Advanced Multilevel Modeling}.
\newblock SAGE Publications Ltd, 2nd edition.

\bibitem[Steenbergen and Jones, 2002]{steenbergen2002}
Steenbergen, M.~R. and Jones, B.~S. (2002).
\newblock Modeling multilevel data structures.
\newblock {\em American Journal of Political Science}, 46(1):218--237.
\newblock \href {https://doi.org/10.2307/3088424} {\path{DOI: 10.2307/3088424}}.

\bibitem[Yarnell et~al., 2019]{yarnell2019}
Yarnell, C., Pinto, R., and Fowler, R. (2019).
\newblock Measuring variability between clusters by subgroup: An extension of the median odds ratio.
\newblock {\em Statistics in Medicine}, 38(22):4253--4263.
\newblock \href {https://doi.org/10.1002/sim.8286} {\path{DOI: 10.1002/sim.8286}}.

\end{thebibliography}
\end{document}